# Path probability density functions for semi-Markovian random walks

O. Flomenbom & R. J. Silbey

*Department of Chemistry, Massachusetts Institute of Technology, Cambridge, MA 02139*

**Abstract**

In random walks, the path representation of the Green's function is an infinite sum over the length of path probability density functions (PDFs). Here we derive and solve, in Laplace space, the recursion relation for the $n$ order path PDF for any *arbitrarily inhomogeneous* semi-Markovian random walk in a one-dimensional (1D) chain of $L$ states. The recursion relation relates the $n$ order path PDF to $L/2$ (round towards zero for an odd $L$) shorter path PDFs, and has $n$ independent coefficients that obey a universal formula. The $z$ transform of the recursion relation straightforwardly gives the generating function for path PDFs, from which we obtain the Green's function of the random walk, and derive an explicit expression for any path PDF of the random walk. These expressions give the most detailed description of arbitrarily inhomogeneous semi-Markovian random walks in 1D.

***I. Introduction*** Random walks [1-16] appear in the description of a wide variety of processes in biology, chemistry and physics [17-33]. Relevant familiar processes include enzymatic activity [3, 16, 20-23], chemical kinetics [1-3] and polymer dynamics [4]. Random walks are ubiquitous because they supply the platform for explaining the stochastic behavior observed in many processes in nature. Probability density functions (PDFs) in time, which are obtained when averaging over stochastic paths, are the mathematical deterministic objects that describe random processes. These objects can be represented by a number of different formulations: the discrete in space master equation [1-3, 5, 8] and the generalized master equation [9-10], the continuous in space and time Fokker Planck equation [4] and its generalizations [15], continuous time random walk, which represents jump processes that are either discrete or continuous in space [6, 9-12], and the renewal theory [7]. For each of these formulations there is also a path representation, e.g. [4, 7, 13-14]. The network of relationships among the various descriptions provides a powerful tool in the analysis of random walks.

Path PDFs are obtained when averaging only over stochastic paths of the same length. Path PDFs analysis is found useful in many systems, e.g. [4-5], and enter naturally in the evolving field of single molecules [17-33]. In single molecules, path PDFs and special correlation functions, which cannot be obtained from ensemble measurements, enable the discrimination of distinct random walk models that have some properties in common, e.g. the bulk relaxation function [19-22, 24-28]. This, in turn, demands a detailed theoretical analysis of random walks. In particular, many of the recent measurements of single molecules yield two-state trajectories [17-22, 24-31]. Path PDFs of two successive events contain all the information in such data [26, 28-29]. There are systems, e.g. ion channels [18] and motor proteins [23], whose measured dynamics are random walks along $L$ (>2) states. Path PDFs can be constructed from such measurements and used to explain them.

In this Letter, we study path PDFs of semi-Markovian random walks in arbitrarily inhomogeneous one-dimensional (1D) chains with $L$ states. A semi-Markovian random walk is a random walk whose dynamics are described by the (possibly) state- and direction-dependent waiting time (WT-) PDFs, $\psi_{i\pm1i}(t) = \omega_{i\pm1i}\varphi_{i\pm1i}(t)$, for transitions between states $i$ and $i\pm1$, that generates stochastic trajectories of uncorrelated waiting times which are non-exponential distributed. $\omega_{kj}$ is the transition probability from state $j$ to state $k$, and obeys the normalization condition, $\sum_k \omega_{kj} = 1 \;\; \forall j$. $\varphi_{kj}(t)$ is a normalized WT-PDF for the transition from state $j$ to state $k$, and obeys $\int_0^\infty \varphi_{kj}(t)dt = 1$, $\forall k$ and $\forall j$ (fig. 1). The dynamics can also include state- and direction-dependent irreversible trapping WT-PDFs, $\psi_{Ii}(t) = \omega_{Ii}\varphi_{Ii}(t)$, with $I=i+L$, and these trapping WT-PDFs determine the boundary conditions. The environment is arbitrarily inhomogeneous when a different WT-PDF is assigned to each transition, and without specifying a functional form to any WT-PDF. (Mathematically, the arbitrarily inhomogeneous environment forces us to track all paths in the calculations of path PDFs. Clearly, any particular case, e.g. the homogeneous case, can be obtained from the results for path PDFs in an arbitrarily inhomogeneous environment.) The path representation of the Green's function $G_{ij}(t;L)$ of the process, which is the PDF of occupying state $i$ at time $t$ when starting at state $j$ exactly at time $0$, is formally given by,

$$G_{ij}(t;L) = \int_0^t \Psi_i(t-\tau)\left(\sum_{n=0}^\infty w_{ij}(\tau, 2n+\gamma_{ij};L)\right)d\tau. \qquad (1)$$

In eq. (1), $\Psi_j(t) = \sum_k \Psi_{kj}(t)$, where $\overline{\Psi}_{kj}(s) = \omega_{kj}[1-\overline{\varphi}_{kj}(s)]/s$, and $\overline{g}(s) = \int_0^\infty g(t)e^{-st}dt$ is the Laplace transform of $g(t)$. The above process is also a continuous time random walk and has an equivalent generalized master equation representation for the Green's function [9-10, 13-14]. The expression in the large parenthesis in eq. (1) defines $W_{ij}(t;L)$,

$$W_{ij}(t;L) = \sum_{n=0}^\infty w_{ij}(t, 2n+\gamma_{ij};L), \qquad (2)$$

where $w_{ij}(t, 2n+\gamma_{ij};L)$ is the path PDF that is built from all paths with $2n+\gamma_{ij}$ ($\gamma_{ij} = |i-j|$) transitions connecting states $j$ to $i$, and each path lasts exactly time $t$. Two different path types contribute to $w_{ij}(t, 2n+\gamma_{ij};L)$ [14]: (1) paths made of the same states appearing in different orders and (2) different paths of the same length of $2n+\gamma_{ij}$ transitions. Path PDFs for translation invariant chains are mono-peaked in time. Path PDFs for translation invariant chains largely contribute to the Green's function in the vicinity of its peak, and this relationship should hold in inhomogeneous chains as well.

In this Letter, we study path PDFs of the most general random walk in 1D. We derive and solve, in Laplace space, the recursion relation for $\overline{w}_{1L}(s, 2n+\gamma_{1L};L)$ in the length $n$ of path PDFs for any fixed value of $L$. The recursion relation is linear in path PDFs with the $n$ independent coefficients, and is of order $L/2$ (round towards zero for an odd $L$). The recursion relation for path PDFs, given in eq. (10), and its solution, given in eqs. (12)-(15), are the main results of this Letter. The path PDFs of this Letter are useful in the analysis of experimental measurements and in theoretical analysis of random walks. For example, path PDFs are the building blocks in approximations of Green's functions. As shown here, path PDFs are useful in explaining properties of the Green's function.

***II. The Green's function*** For the current discussion, it is important to present the closed-form expression for $G_{ij}(t;L)$ in eq. (1) for a random walk in an *L*-state arbitrarily inhomogeneous 1D chain. This solution was obtained recently in [13], by utilizing the path representation of the Green's function. However, a recursion relation for a general order path PDF was not derived in [13], and consequently the expression for the *n* order path PDF was not found. As noted in the introduction, the analysis of path PDFs given in this Letter is used to explain some properties of the Green's function that is given below in eqs. (3)-(6).

The formula for the Green's function for the most general 1D random walk, given in terms of the input WT-PDFs, reads [13],

$$\overline{G}_{ij}(s;L) = \overline{\Gamma}_{ij}(s) \frac{\overline{\Phi}(s;\widetilde{L})}{\overline{\Phi}(s;L)} \overline{\Psi}_i(s) \equiv \overline{W}_{ij}(s;L)\overline{\Psi}_i(s), \tag{3}$$

where $\widetilde{L} = L - \gamma_{ij}$. In eq. (3),

$$\overline{\Gamma}_{ij}(s) = \prod_{k=j}^{i\mp 1} \overline{\psi}_{k\pm 1 k}(s), \; i \neq j \; ; \qquad \overline{\Gamma}_{ii}(s) = 1. \tag{4}$$

$\overline{\Gamma}_{ij}(s)$ is the path of direct transitions connecting the initial and final states. (The plus (minus) sign in eq. (4) corresponds to the case $i > j$ ($i < j$).) The factor $\overline{\Phi}(s;\widetilde{L})/\overline{\Phi}(s;L)$ originates from all possible transitions between the initial and final states. $\overline{\Phi}(s;L)$ depends only on the system length *L*,

$$\overline{\Phi}(s;L) = 1 + \sum_{i=1}^{[L/2]} (-1)^i \overline{h}(s,i;L) \quad ; \quad L>1, \tag{5}$$

with,

$$\overline{h}(s,i;L) = \prod_{j=1}^{i} \sum_{k_j = k_{j-1}+2}^{L-1-2(i-j)} \overline{f}_{k_j}(s) \quad ; \quad \overline{f}_{k_j}(s) = \overline{\psi}_{k_j k_j+1}(s)\overline{\psi}_{k_j+1 k_j}(s), \tag{6}$$

where $k_0 = -1$. For $L=1$, $\overline{\Phi}(s;1) \equiv 1$. In this Letter, the symbol $[L/2]$, as appearing in the upper bound of the sum in eq. (6), is the floor operation (round towards zero). Finally, the factor $\overline{\Phi}(s;\widetilde{L})$ in eq. (3) has the same form as $\overline{\Phi}(s;L)$, given by eqs. (5)-(6), but it is calculated on a lattice $\widetilde{L}$. Lattice $\widetilde{L}$ is constructed from the original lattice by removing the states $i$ and $j$ and the states between them, and then connecting the obtained two fragments. The states in each of the fragments have their corresponding WT-PDFs. The WT-PDFs in the interface between the two fragments do not exist in the original lattice, and, therefore, these WT-PDFs vanish in lattice $\widetilde{L}$. For cases in which a fragment is a single state, this fragment is excluded; namely, lattice $\widetilde{L}$ is the longer fragment. When each fragment is a single state, $\overline{\Phi}(s;\widetilde{L}) = 1$.

Clearly, $\overline{G}_{ij}(s;L)$ in eqs. (3)-(6) solves the corresponding continuous time random walk problem and the equivalent generalized master equation. Equations (3)-(6) enable analyzing semi-Markovian random walks in 1D chains from a wide variety of aspects. Inversion to time domain gives the Green's function, but also moments and correlation functions can be calculated from eqs. (3)-(6), and then inverted into time domain. The closed-form $\overline{G}_{ij}(s;L)$ also manifests its utility when numerical inversion of the generalized master equation is unstable. Moreover, using $\overline{G}_{ij}(s;L)$ in simple analytical manipulations gives [13-14]: (*i*) the first passage time PDF, (*ii*)-(*iii*) the Green's functions for a random walk with a special WT-PDF for the first event and for a random walk in a circular $L$-state 1D chain, and (*iv*) joint PDFs in space and time with many arguments.

***III. Path PDFs*** Complementary information on the random walk to that supplied by the Green's function is contained in path PDFs. Formally, this is evident in eq. (1): path

PDFs partition the Green's function into physical meaningful objects, but knowing the Green's function doesn't give path PDFs. The usefulness stemming from this partition appears in both analytical and numerical studies of random walks and in analysis of experimental data.

We start the analysis of path PDFs by introducing the recursion relations for small chains, which are found by *direct counting* of paths. We then extend these results for a general $L$ using physical and mathematical arguments. We note here that any recursion relation given in this Letter corresponds for path PDFs that connect the edge states, $w_{1L}(t, 2n + \gamma_{1L}; L)$. This is sufficient because $\overline{W}_{ij}(s; L)$, which is the Laplace transform of eq.(2), can be written as a product of two PDFs that connect the edge states of different lattices, $\overline{W}_{ij}(s; L) = \overline{W}_{1L}(s; L) / \overline{W}_{1\widetilde{L}}(s; \widetilde{L})$. Here, $\overline{W}_{1\widetilde{L}}(s; \widetilde{L})$ connects the edge states of the reduced lattice $\widetilde{L}$. Now, recall that $\overline{W}_{ij}(s; L)$ is the $z=1$ point of the $z$ transform ($\hat{g}(z) = \sum_n g_n z^n$) of path PDFs, $\hat{\overline{w}}_{ij}(s, 2z + \gamma_{ij}; L) = \sum_{n=0}^{\infty} \overline{w}_{ij}(s, 2n + \gamma_{ij}; L) z^n$. Continuation of the ratio relationship for $\hat{\overline{w}}_{ij}(s, 2z + \gamma_{ij}; L)\big|_{z=1}$ to any value of $z$ relates path PDFs that connect *internal* states in the original lattice with path PDFs that connect the *edge* states in the original and the reduced lattices.

The recursion relation for path PDFs for a chain of two states contains only one successive shorter path PDF,

$$\overline{w}_{12}(s, 2n+1; 2) = \overline{w}_{12}(s, 2(n-1)+1; 2)\overline{h}(s,1;2) + \delta_{n0}\overline{\Gamma}_{12}(s), \tag{7}$$

and, in fact, $\overline{w}_{12}(s, 2n+1; 2)$ is associated with only one path. (In eq. (7), $\overline{h}(s,1;2) = \overline{f}_1(s)$, as can be obtained from eq. (6) with the appropriate substitution for $L$. However, $\overline{h}(s,1;2)$ in eq. (7) is found by a direct counting of paths and is *not* a quote of eq. (6). This statement is true for all the $\bar{h}(s,i;L)$ s given in the recursion relations for particular chain lengths derived in this Letter, i.e. eqs. (7)-(9).) The second term in the right hand side (RHS) in eq. (7) represents the initial condition. Note that the notation in the path-length argument of $\bar{w}_{12}(s,2(n-1)+1;2)$ distinguishes between the contribution to the path PDF from the back transitions, $2(n-1)$, and the contribution to the path from the direct transitions, $1 = \gamma_{12}$. Also, for $n$ values that lead to $(n-i)<0$ in $\bar{w}_{12}(s,2(n-i)+\gamma_{12};2)$, the path PDF is set to zero. These notation and convention are also used in path PDFs of larger chains.

For $L=3$, the recursion relation for $\bar{w}_{13}(s,2n+2;3)$ is formally identical to the $L=2$ case,

$$\bar{w}_{13}(s,2n+2;3) = \bar{w}_{13}(s,2(n-1)+2;3)\bar{h}(s,1;3) + \delta_{n0}\bar{\Gamma}_{13}(s),  \tag{8}$$

Equations (7)-(8) are formally identical and this means that path PDFs for $L=2$ and $L=3$ are also formally identical. Specifically, $\bar{w}_{1L}(s,2n+\gamma_{1L};L) = \bar{\Gamma}_{1L}(s)\left(\bar{h}(s,1;L)\right)^n$ for $L=2, 3$. The important difference between $\bar{w}_{12}(s,2n+1;2)$ and $\bar{w}_{13}(s,2n+2;3)$ lies in the different forms of the $\bar{h}(s,1;L)$, which is also the important difference in the corresponding recursion relations: $\bar{h}(s,1;3) = \bar{f}_1(s) + \bar{f}_2(s)$, but $\bar{h}(s,1;2) = \bar{f}_1(s)$. The fact that $\bar{h}(s,1;3)$ is a sum of two terms indicates that $\bar{w}_{13}(s,2n+2;3)$ originates from more than one path. In fact, $\bar{w}_{13}(s,2n+2;3)$ originates from $2^n$ paths [14].

Next, we consider chains of length $L=4$ and $L=5$, for which the recursion relations are also formally identical to each other, but, in contrast to the $L=2, 3$ cases, require information from two successive shorter path PDFs. Moreover, the shorter path PDF appearing in the recursion relation has a negative sign, which means subtraction of paths. The recursion relation for $\overline{w}_{1L}(s,2n+\gamma_{1L};L)$, $L=4, 5$, reads,

$$\overline{w}_{1L}(s,2n+\gamma_{1L};L) = \overline{w}_{1L}(s,2(n-1)+\gamma_{1L};L)\overline{h}(s,1;L)$$

$$-\overline{w}_{1L}(s,2(n-2)+\gamma_{1L};L)\overline{h}(s,2;L) + \delta_{n0}\overline{\Gamma}_{1L}(s). \qquad (9)$$

If the second term on the RHS in eq. (9) is neglected, the $n$ order path PDF is simply the $n$ power of $\overline{h}(s,1;L)$, as is observed for the smaller chains. This scaling, however, represents also discontinuous paths [13-14]. The correction is proportional to the path PDF of order $n$-2. The proportionality constant must have the length of four transitions, by demanding conservation of path length. $\overline{h}(s,2;L)$ satisfies this demand.

Based on the derived recursion relations for small chains, we introduce the recursion relation for $\overline{w}_{1L}(s,2n+\gamma_{1L};L)$ for a general $L$,

$$\overline{w}_{1L}(s,2n+\gamma_{1L};L) = \sum_{i=1}^{[L/2]}(-1)^{i+1}\overline{h}(s,i;L)\overline{w}_{1L}(s,2(n-i)+\gamma_{1L};L) + \delta_{n0}\overline{\Gamma}_{1L}(s). \qquad (10)$$

Equation (10) has a convolution form both in path length and in time. An educated guess could suggest a convolution form for such a quantity. The calculations for the smaller chains not only verify this form, but also enable the determination of the details of the recursion relation for $\overline{w}_{1L}(s,2n+\gamma_{1L};L)$ for a general $L$. The non-trivial details in eq. (10) are the order of the recursion relation and its coefficients.

We explain eq. (10) by the following reasoning. The recursion relations for the small chains are linear in path PDFs with $n$ independent coefficients, and these two properties *must* be independent of $L$. The same is true for the order of the recursion relation; it was shown that the recursion relation for a chain of $L$ =2, 3, 4, 5, states is of the order of $[L/2]$, and this property must be independent of the specific value of $L$. Equation (10) leads to the Green's function in eqs. (3)–(6) when the coefficients in the recursion relation are the $\bar{h}(s,i;L)$s of eq. (6). This is seen by applying a $z$ transform on eq. (10). Utilizing the fact that the recursion relation is linear in path PDFs with $n$-independent coefficients, the $z$ transform is easily done, and leads to the expression for the path PDF generating function,

$$\hat{\overline{w}}_{1L}(s,2z+\gamma_{1L};L) = \sum_{n=0}^{\infty} \overline{w}_{1L}(s,2n+\gamma_{1L};L)z^n = \overline{\Gamma}_{1L}(s)\left(1 - \sum_{i=1}^{[L/2]}(-1)^{i+1}\bar{h}(s,i;L)z^i\right)^{-1}. \quad (11)$$

Setting $z=1$ in eq. (11) leads to $\overline{W}_{1L}(s;L)$. Taking the $n$ derivative (with respect to $z$) of eq. (11) and substituting $z=0$ gives $n!\overline{w}_{1L}(s,2n+\gamma_{1L};L)$, but $\overline{w}_{1L}(s,2n+\gamma_{1L};L)$ can also be obtained from a Taylor expansion of eq. (11). The result can be written in the form,

$$\overline{w}_{1L}(s,2n+\gamma_{1L};L) = \overline{\Gamma}_{1L}(s)\sum_{k_0=a_{0,n}}^{n}\left(\bar{h}(1,s;L)\right)^{k_0}\overline{c}_{k_0}(s;L), \quad (12)$$

where $\overline{c}_{k_0}(s;L)$ is given by,

$$\overline{c}_{k_0}(s;L) = \prod_{i=1}^{[L/2]-1}\sum_{k_i=a_{i,n}}^{n-\sum_{j=0}^{i-1}k_j}\overline{g}_{k_i}(s;L), \quad (13)$$

with,

$$\overline{g}_{k_i}(s;L) = \binom{k_{i-1}}{k_i}\left(-\frac{\bar{h}(s,i+1;L)}{\bar{h}(s,i;L)}\right)^{k_i}. \quad (14)$$

For $L=2, 3$, $\overline{c}_{k_0}(s;L) \equiv 1$. In eqs. (12)-(13), the parameter $a_{i,n}$ that appears in lower bound of summations is given by,

$$a_{i,n} = \left[\frac{n - \sum_{j=0}^{i-1}k_j + [L/2] - 1 - i}{[L/2] - i}\right] \quad ; \quad i > 0, \quad (15)$$

and $a_{0,n} = \left[\dfrac{n+[L/2]-1}{[L/2]}\right]$.

*IV. Concluding remarks* In this Letter, we studied path PDFs of the most general random walk in 1D. We derived and solved a recursion relation for the path PDFs. The formula for the *n* order path PDF, given in eqs. (12)-(15), can be easily implemented in a computer program, but can also be a starting point in an analytical analysis. In an invariant chain, $\overline{w}_{1L}(s, 2n + \gamma_{1L}; L)$ can be approximated by, $\overline{w}_{1L}(s, 2n + \gamma_{1L}; L) = 2P_L(ss)(\overline{\varphi}(s))^{L-1+2n}$, where $P_L(ss)$ is the probability to occupy state *L* in steady state [14]. The recursion relation for path PDF in eq. (10), its relationship to the Green's function in eqs. (3)-(6) (see also the remark below), and the general expression for path PDFs in eqs. (12)-(15) are the main results of this Letter.

The final point in this Letter gives support to the universal formula for $\overline{h}(s,i;L)$ in eq. (6). (The term 'universal formula' means that one formula defines $\overline{h}(s,i;L)$ for a fixed *i* and any *L*, and for a fixed *L* and any *i*.) Firstly, recall that the structure of the recursion relation in eq. (10) is universal, and this indicates that the coefficients should also be universal. Now, $\overline{h}(s,1;L)$ is universal because $\overline{h}(s,1;L)\overline{\Gamma}_{1L}(s)$ is the path PDF that contains all paths with *one* back transition, so, $\overline{h}(s,1;L) = \sum_{i=1}^{L-1} \overline{f}_i(s)$ for any *L*. We extend this universality of $\overline{h}(s,i;L)$ with *L* and any fixed *i* by noting that when a given $\overline{h}(s,i;L)$ first enters the recursion relation, the recursion relation has the same form as the recursion relations for all the smaller chains that contain the *i* order. (This is a consequence of the demand that for values of *n* that lead to (*n-i*)<0 in $\overline{w}_{1L}(s, 2(n-i) + \gamma_{1L}; L)$ in a RHS of a recursion relation, the path PDF is set to zero.) The equivalence of recursion relations for a fix *n* ($<[L/2]$) as a function of *L* means that the formal expressions for the *L* dependent path PDFs for a fix *n* ($<[L/2]$) are the same.

Thus, the correction term $(-1)^{i+1}\bar{h}(s,i;L)\bar{w}_{1L}(s,2(n-i)+\gamma_{1L};L)$ as a function of $L$ for any fixed $i$ originates from the same physical reasons (counting of paths with the same length of back transitions). Together, these three points justify the universality of the coefficient $\bar{h}(s,i;L)$ with $L$ for *any* fixed $i$. Technically, eq. (6) is derived by recasting into one equation the form of $\bar{h}(s,i;L)$ calculated for $i=1, 2, 3$, for chain lengths obeying $2 \le L \le 6$. (For $L=6, 7$, direct counting straightforwardly gives, $\bar{h}(s,3;L) = \sum_{i=1}^{L-5} \bar{f}_i(s) \sum_{j=i+2}^{L-3} \bar{f}_j(s) \sum_{k=j+2}^{L-1} \bar{f}_k(s)$.) Cleary, a numerical verification of the form of the Green's function in eqs. (3)-(6) can be easily done for the Markovian case for any reasonable value $L$, and can supply an indirect verification of the formula for $\bar{h}(s,i;L)$. In addition, a direct verification of the form of the $\bar{h}(s,i;L)$s for any $i > 3$ can be done analytically by constructing eq. (10) for any fixed relevant value of $L$.

**Figure captions**

**FIG 1** A part of a semi-Markovian chain in 1D with directional WT-PDFs, $\psi_{i\pm 1 i}(t) = \omega_{i\pm 1 i}\varphi_{i\pm 1 i}(t)$ and $\psi_{Ii}(t) = \omega_{Ii}\varphi_{Ii}(t)$. A way to simulate such a random walk is to first draw a random number out of a uniform distribution that determines the propagation direction according to the transition probabilities, and then to draw a random time out of the relevant WT-PDF.

**FIGURE 1**

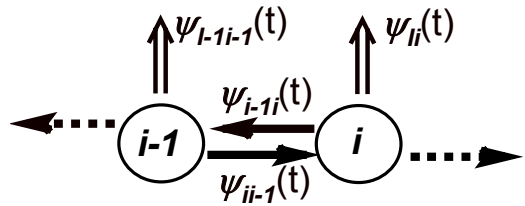